\documentstyle[12pt]{article}
\begin{document}
\begin{titlepage}
\vspace{.3in}
\begin{center}
{\Large{\bf Improving the
Convergence of SU(3) Baryon Chiral Perturbation Theory }}\\
\vskip 0.3in
{\bf John F. Donoghue$^1$ and Barry R. Holstein$^{1,2}$}\\[.2in]
{\it $^1$ Department of Physics and Astronomy, \\
University of Massachusetts,
Amherst, MA 01003 \\
$^2$ Insititut f\"{u}r Kernphyisk\\
Forschungszemtrum J\"{u}lich\\
D-52425 J\"{u}lich, Germany  }

\end{center}
\vskip 0.4in

\begin{abstract}
Baryon chiral perturbation theory as conventionally applied using
dimensional regularization has a
well-known problem with the convergence of the SU(3)
chiral expansion. One can
reformulate the theory equally rigorously using a momentum-space
cutoff and we show that the convergence is thereby greatly improved for
reasonable values of the cutoff. In effect, this is accomplished
because the cutoff formalism  removes the
spurious physics of propagation at distances
much smaller than the baryon size.
\end{abstract}
\end{titlepage}

\section{Basic ideas}

Chiral perturbation theory describes low energy QCD via a simultaneous
expansion in the energy and in quark masses.\cite{1,2} The original application
to the physics of pions and kaons has enjoyed a remarkable success and
the convergence of the energy expansion has been acceptable for
realistic applications. The formalism in the baryon sector has also
been developed,\cite{66,67,68,69,70},
but here one finds problems with the convergence of the
chiral expansion, especially in the application of chiral
SU(3)---typically, SU(3) symmetry relations are well satisfied at tree-level
(i.e. to first order in the quark masses), but when ``improved'' to
one-loop the corrections are larger than are allowed by experiment.
Well known examples include

\begin{itemize}
\item [i)] Baryon masses, wherein the Gell-Mann-Okubo relation obtains
at first order in quark mass and is well satisfield experimentally.
Chiral loops, however, make significant---{\cal
O}(50-100\%)---corrections to individual masses.\cite{3}

\item [ii)] Semileptonic hyperon decay, wherein a simple SU(3)
representation of the axial couplings in terms of F,D couplings yields
an excellent fit to experiment.  Chiral loops make {\cal O}(30-50\%)
corrections to individual couplings and destroy this agreement.\cite{4}

\item [iii)] Nonleptonic hyperon decay, wherein a simple SU(3) fit to
s-wave amplitudes provides an excellent representation of the
experimental numbers in terms of f,d couplings.  Chiral loops make
{\cal O}(30-50\%) corrections to individual terms and destroy this
agreement.  The situation is somewhat more confused in the case of the
p-waves wherein a significant cancellation between pole terms exists
at lowest order and the validity of the chiral expansion is suspect.\cite{5}

\end{itemize}

\noindent Of course, these loop modifications can in general
be rescued by including yet higher orders in the chiral expansion, but
the net result is that the expansion is not well behaved in that there is
no clear convergence to the orders yet calculated.

In this paper, we show that this SU(3) violation from
loop diagrams arises to a large extent from propagation at short
distances---smaller than the physical size of baryons---where the
effective field theory cannot properly represent the correct physics.
If we remove this short-distance physics by the application of a
cutoff, we obtain  a greatly improved phenomenology and a nicely
convergent expansion. Chiral perturbation theory can be formulated
equally rigorously with a cutoff as with the conventionally employed
dimensional regularization,\cite{7} and we will demonstrate how this works in
a specific example below. We shall show that the formalism
of baryon chiral perturbation theory with a cutoff then
opens the possibility for useful SU(3) chiral phenomenology in the baryon
sector.

Effective field theory is a technique which uses the very low energy
interactions and degrees of freedom of a theory in order to calculate the long
distance physics appropriate for low energy problems.\cite{6} The
effects of short-distances/high-energies are not calculated directly
but are encoded in the
coefficients of a general local effective Lagrangian. When loop
diagrams are calculated, the reliable portion of the result comes only
from the long-distance/low-energy portion of the loop, as that is the
only portion for which the effective theory is appropriate. In the
same diagrams, there are always also contributions from high energy
which are not correctly represented by the effective theory. However,
this is not a problem in principle because these spurious high energy
contributions are equivalent to terms in the local effective
Lagrangian and can be corrected for by a shift in coefficients of this
Lagrangian. The formalism then allows one to calculate rigorously the
long distance dynamics of loop processes while providing a general
parameterization of the short distance physics.

In the interaction of pions with baryons, the long distance physics is
the propagation of pions out to large distances---the so-called ``pion
tail''. Because the pion is light its propagator has
significant strength at distances of order of its Compton wavelength,
{\it i.e.} $\sim$1.4 fermis. The effective field theory formalism
represents baryons and pions as point particles, even though we know
that they  have a non-zero size. This is not a problem for the
lowest energy propagation---the pion tail is model independent.
However, this
feature tells us where ``short distance'' physics starts. The
effective theory in terms of point particles misrepresents the physics on
distance scales smaller than the actual size of the particles.  For
example, a baryon has a
rms charge radius of $\sim$0.8 fermis, so that loop effects below
this distance are not reliably calculable in the effective theory of
point baryons.

Chiral loops do in fact generate large contributions from short
distance---in general they are divergent.  The most common
regularization scheme to avoid this problem
is dimensional regularization, which is elegant
and simple to apply. In such a scheme, in addition to the correct long
distance behavior, there is a residual dependence on the short distance
portion of the loop integrals. It is also possible to regularize the loop
integrals by a momentum-space cutoff.\cite{7} This effectively removes
the short distance propagation. Either scheme can be used {\it equally
rigorously} in a chiral effective field theory, as long as care is
taken to preserve the chiral symmetry. Since the schemes differ only
in the treatment of the short distance portion of the theory, they
will involve different coefficients of higher order terms in the
chiral expansion. However, as long as these are treated fully
generally, the same physics must result.

We will demonstrate below how the use of a cutoff can be applied in baryon
chiral perturbation theory and how it resolves the problem with
the convergence of the energy expansion. A longer paper will present
full details and additional examples of this reformulation of the
theory,\cite{8} while here we
concentrate simply on the underlying physics. First we demonstrate how to
reproduce known results in baryon masses using the cutoff formalism.
This includes several non-trivial consistancy checks. Subsequently, by
chosing a realistic value of the cutoff, representing the onset of
short distance physics, we show how the problematic (and apparently
unphysical) symmetry breaking effects generated by loop contributions
are moderated in this formalism.

\section{Baryon masses}

An example of the difficulties of baryon chiral perturbation
theory is provided by the analysis of baryon masses. The masses
have an expansion in the masses of the quarks ($m_{q}$),
or equivalently in terms of the pseudoscalar meson masses ($m_{M}$)
\begin{eqnarray}
M_B & = & M_0 + \Sigma_q ~\bar{b}_q m_{q} + \Sigma_q ~\bar{c}_q m_{q}^{3/2}
 + \Sigma_q ~\bar{d}_q m_{q}^2 + ...  \\
 & = & M_0 + \Sigma_M ~{b}_M m_{M}^2 + \Sigma_M ~{c}_M m_{M}^{3}
 + \Sigma_M ~{d}_M m_{M}^4 + ...
\end{eqnarray}
\noindent Here $M_0$ is a common mass and $b_M$ and $d_M$ contain adustable
parameters representing terms in the effective Lagrangian. However,
the non-analytic $m_q^{3/2}$ terms come from loop diagrams, and
the coefficients are not adjustable but are known in terms of
the baryon-meson coupling constants. The leading SU(3) breaking
terms involving $b_M$ go back to Gell-Mann and Okubo\cite{9}, the non-analytic
corrections from one-loop diagrams, represented above by $c_M$,
were first calculated by Langacker and Pagels\cite{10}, and
$m_M^4$ corrections (including diagrams up to two loops) were
calculated by Borasoy and Meissner\cite{3}. The convergence difficulties
in the expansion are demonstrated by the resulting fit for the
nucleon mass where, in the same sequence, the different
contributions are given, in GeV, by\cite{3}
\begin{equation}
M_N = 0.711 + 0.202 - 0.272 + 0.298+\ldots
\end{equation}
or, more dramatically for the $\Xi$,
\begin{equation}
M_\Xi = 0.767 +0.844  - 0.890 + 0.600+\ldots
\end{equation}
\noindent The non-analytic terms appear unavoidably large and the
expansion has certainly not converged at this order. The
final fit also violates the Gell-Mann-Okubo relation by an amount that
is five times larger than the experimentally observed violation.

To one-loop order, the explicit form of the contributions to
the baryon masses is given by\cite{70}
\begin{eqnarray}
M_N & = & {\hat M}_0 - 4m_K^2 b_D + 4(m_K^2 - m_\pi^2) b_F + L_N  \nonumber\\
M_\Lambda & = &  {\hat M}_0-{4 \over 3}(4 m_K^2 - m_\pi^2) b_D +
L_\Lambda \nonumber\\
M_\Sigma & = &  {\hat M}_0 - 4 m_\pi^2 b_D + L_\Sigma \nonumber\\
M_\Xi & = &  {\hat M}_0 - 4m_K^2 b_D - 4(m_K^2 - m_\pi^2) b_F + L_\Xi
\end{eqnarray}
where
\begin{equation}
 {\hat M}_0 = M_0 - 2(2 m_K^2 + m_\pi^2) b_0
\end{equation}
with $M_0$, $b_D$, $b_F$ and $b_0$ as free parameters. (Note that
$M_0$ and $b_0$ do not have separate effects, but only enter in the
combination ${\hat M}_0$.) The ingredients $L_B$ contain the
nonanalytic contributions from loop diagrams, and have the form\cite{11}
\begin{eqnarray}
L_N & = & -
{1 \over 24 \pi F_\pi^2} \left[ {9 \over 4}(D+F)^2 m_\pi^3
+ {1 \over 2}(5 D^2 - 6DF +9F^2) m_K^3 + {1 \over 4} (D-3F)^2 m_\eta^3
\right] \nonumber\\
L_\Lambda & = &-{1 \over 24 \pi F_\pi^2} \left[ 3D^2 m_\pi^3
+ (D^2 +9F^2) m_K^3 + D^2 m_\eta^3
\right] \nonumber\\
L_\Sigma & = &-{1 \over 24 \pi F_\pi^2} \left[ (D^2+6F^2) m_\pi^3
+ 3 (D^2  +F^2) m_K^3 +  D^2 m_\eta^3
\right] \nonumber\\
L_\Xi & = &-{1 \over 24 \pi F_\pi^2} \left[ {9 \over 4}(D-F)^2 m_\pi^3
+ {1 \over 2}(5 D^2 + 6DF +9F^2) m_K^3 + {1 \over 4} (D+3F)^2 m_\eta^3
\right] \nonumber\\
\qquad\label{eq:bb}
\end{eqnarray}
where $D$ and $F$ parameterize the baryon axial-vextor current ($D+F =
1.266$ and $D/(D+F) = 0.64$).
The non-analytic terms are quite large, having values
\begin{eqnarray}
L_N & = & -0.31 ~GeV, \nonumber\\
L_\Lambda & = & -0.66 ~GeV, \nonumber\\
L_\Sigma & = & -0.67 ~GeV, \nonumber\\
L_\Xi & = & -1.02 ~GeV.
\end{eqnarray}
using $D = 0.806$ and $F = 0.46$ and $F_\pi = 93$MeV.
(The slight numerical disagreement
with the fit quoted in Eq. 3,4 occurs because the authors of Ref. 3 used
somewhat different D,F and $F_\pi$ values.)
In particular, the $\Xi$ mass shift is
clearly unphysically large.
It is not possible to obtain a reasonably convergent
fit to the masses with these large non-analytic terms to this order.

\section{Regularization with a cutoff}

The mass analysis is especially simple in the heavy baryon
formalism\cite{67, 68, 69}
using dimensional regularization where all of the mass
shifts are proportional to a single integral
\begin{equation}
I(m_P^2) = {1 \over i}
\int {d^4k \over (2\pi)^4} {{\bf k \cdot k} \over (k_0 - i \epsilon)
(k^2 - m_P^2+i\epsilon ) }\label{eq:aa}
\end{equation}
where $m_P$ is the mass of the Goldstone boson that is involved in the loop.
When dimensionally regularized, this has results in
\begin{equation}
I(m_P^2) = {m_P^3 \over 8 \pi}
\end{equation}
Observe that a peculiarity of dimensional regularization is that,
although this integral appear cubicly divergent via power-counting,
the dimensionally regularized form
is finite even for $d \rightarrow 4$.
There is also the counter-intuitive feature that the result vanishes
in the massless limit, wherein we would expect the long distance physics
to be the most important, and also grows larger for the more massive
states than for the pion, while physically we would expect the
reverse. It appears that a short-distance subtraction is implicit
in this formalism.
Since the meson mass-squared is proportional
to the quark mass, this integral is uniquely the source of the
non-analytic terms in the previous results, $L_B$.

Let us now calculate this integral with a momentum space cutoff.\cite{11}
There are many forms which could be employed equivalently, and
we chose one possibility with an eye to the finite spatial size
of the baryons. Since the heavy baryon defines a preferred frame of
reference---the rest frame---we may choose a form which regulates
only the spatial momentum components in that frame. In covariant
notation this is, {\it e.g.}, a cutoff of the form
\begin{equation}
e^{k^2 - (v\cdot k)^2 \over \Lambda^2} \sim e^{-{{\bf k}^2 \over
\Lambda^2}}
\end{equation}
where $v_\mu$ is the unit four-vector that defines the rest frame
of the baryon. With this regularization the integral becomes
\begin{equation}
   I(m_P^2) = {1 \over 8 \pi} \left[ {\Lambda^3 \over 2 \sqrt{\pi}} -
{ \Lambda m_P^2 \over \sqrt{\pi}} + m_P^3 e^{m_P^2 \over \Lambda^2}
\left( 1 - \Phi({m_P \over \Lambda}) \right) \right]
\end{equation}
\noindent where $\Phi(x)$ is the probability integral
\begin{equation}
\Phi (x) = {2 \over \sqrt{\pi}} \int_0^x dt~ e^{-t^2} \ \ .
\end{equation}
When the meson mass is small compared to the cutoff, the integral
simplifies to
\begin{equation}
   I(m_P^2)\stackrel{m_P^2<<\Lambda^2}{\longrightarrow}
{1 \over 8 \pi} \left[ {\Lambda^3 \over 2 \sqrt{\pi}} -
{ \Lambda m_P^2 \over \sqrt{\pi}} + {m_P^3 } \right]
\end{equation}

Let us demonstrate how the usual results are recovered in this
limit. The $\Lambda^3$ contribution yields simply an overall shift to
the baryon mass, of the same form as $M_0$, while the $\Lambda m_P^2$
contribution has the same form as the first SU(3) breaking terms.
Thus the results at this order have the generic form
\begin{equation}
M_B = (M_0 + k \Lambda^3) + \sum (b_P - k \Lambda) m_{P}^2
 +\sum c_P m_{P}^{3}
\end{equation}
The dependence on $\Lambda$ can be completely absorbed into
renormalized values of $M_0$ and $b_i$. To accomplish this
in practice in fact requires highly non-trivial consistency
checks. There are four baryon masses, and each mass expression
contains the loop integral for pions, kaons and etas. However,
there is only a single $M_0$, and three $b_i$ that must absorb
all dependence on $\Lambda$. Thus there are numerous consistency
requirements if this is to work properly.  One can check that all are
satisfied and that all
$\Lambda $ dependence is completely absorbed via the definitions
\begin{eqnarray}
M_0^{ren} & = & M_0 - {5D^2 + 9 F^2 \over 24\pi F_\pi^2} \Lambda^3
    \nonumber\\
b_D^{ren} & = & b_D + {3F^2 - D^2 \over 64\pi F_\pi^2} \Lambda    \nonumber\\
b_F^{ren} & = & b_F + {5DF \over 96\pi F_\pi^2} \Lambda   \nonumber\\
b_0^{ren} & = & b_0 + {13D^2 + 9F^2 \over 288\pi F_\pi^2} \Lambda
\end{eqnarray}
This is a verification that our cutoff regularization respects the chiral
cymmetry, as expected. Thus when
the meson masses are small compared
to the cutoff, the usual analysis is completely reproduced.

On the other hand, when the meson mass
becomes large compared to the cutoff, the effect of the loop
diagram is moderated. For example, if the mass is large compared
to the cutoff, we have the result
\begin{equation}
 I(m_P^2) = - {3  \over 32 \pi^{3/2}} {\Lambda^5 \over m_P^2} \ \ ,
\end{equation}
i.e. it vanishes for large $m_P$.
This is physically reasonable, as for large mass only a vanishingly
small portion of the loop integral occurs at low momentum. Note
that the dimensionally regularized form of the integral does not
allow this distinction---the result grows
continuously with increasing
meson mass and there is no separation of the short and long distance
components.

When employing any regularization scheme that introduces a dimesionful
parameter, the usual power counting rules will be upset. This is
manifest in the results quoted above, in which the lowest order chiral
parameter, $M_0$ is shifted by the loop correction. However, since
these shifts are just the renormalization of phenomenological
parameters, they do
not influence the physics, and a proper chiral expansion of the final results
will always be obtained. In practice, renormalization with a
cut-off is not much more difficult than that in the framework of
dimensional regularization.

Our results do not depend on the specific form of the
cutoff function employed.  For example, we have also
considered a dipole cutoff
\begin{equation}
\left({\Lambda^2\over \Lambda^2-k^2}\right)^2,
\end{equation}
which produces the same qualtitative features with the
replacement
\begin{equation}
I(m_P^2) = {1\over 8\pi}\left[{\Lambda^4\over
(\Lambda^2-m_P^2)^2}(m_P^3-\Lambda^3)
+{3\over 2}{\Lambda^5\over \Lambda^2-m_P^2}\right]
\end{equation}
and the renormalization of the chiral parameters is effected in the
same way.

\section{Phenomenology}

   The formalism may be applied with any value of the cutoff, as long
as the cutoff
 is not chosen so small that it removes physics that is truly
long distance. However, if the cutoff is chosen too large, the
convergence may be poor because the loop
calculation will include spurious short-distance physics which will
have to be removed by counterterms at higher order in the
energy expansion. The best values of the cutoff are then those
close to the scale where the effective field theory description
starts to be inaccurate.

In baryons, the physical size of the hadrons is $\sim$1 fermi. For
propagation over distances much below this scale, the effective field theory
description in terms of point baryons and pions will no longer be
accurate. Therefore we want to choose a cutoff representative of that
scale. In the case of the electromagnetic interaction,
the baryon-photon vertex is known
to have a dipole shape with a mass scale around the rho mass. The
self energy
diagram involves two vertices, so that mimicking the effects of
form-factors would suggest a quartic shape, leading to an estimate of
$\Lambda\sim m_\rho / 2$. Note, however, that strictly speaking we are
not doing field theory with form-factors, but are merely regularizing the
loop integrals with a cutoff. In practice, any cutoff around this
value will be sufficient.

We note that the kaon and eta masses are not sufficiently small
compared to the cutoff that all of their effects need be
long-distance. Employing a reasonable cutoff will keep only the long
distance portion of the loop diagram. We then may use the loop
integral directly in the phenomenology.
For the baryon masses, this leaves the previous formulas of Eq. \ref{eq:bb}
unchanged except for the substitution in the non-analytic terms of the
form
\begin{equation}
m_P^3 \to {\bar I}(m_P^2)
\end{equation}
with
\begin{equation}
{\bar I}(m_P^2) =  m_P^3 e^{m_P^2 \over \Lambda^2}
\left( 1 - \Phi({m_P \over \Lambda}) \right) + {\Lambda^3 \over 2 \sqrt{\pi}} -
{ \Lambda m_P^2 \over \sqrt{\pi}}
\end{equation}
In Table 1, we show that this substitution
significantly moderates the magnitude of
the non-analytic terms for kaons and etas for any reasonable value of
the cutoff. This by itself is a
demonstration that much of the kaon and eta loop integrals, when
calculated dimensionaly, actually correspond to short-distance physics
and are not reliable parts of the chiral effective field theory.
As is physically reasonable, the pion has the largest long-distance
loop effect, and the value decreases as the meson becomes more
massive.
Note that in this instance we are not absorbing the $\Lambda$ into the
chiral parameters, but are keeping the cutoff in the full loop
effect, ${\bar I}(m_P^2)$.

\begin{table}
\begin{center}
\begin{tabular}{c|c|c|c|c}
 &$\Lambda=300$&$\Lambda=400$&$\Lambda=500$&$\Lambda=600$\\
\hline\\
${\bar I}_\pi$&0.00611&0.0157&0.03198&0.0567\\
${\bar I}_K$&0.0024&0.0077&0.01833&0.0363\\
${\bar I}_\eta$&0.0020&0.0069&0.01633&0.0329
\end{tabular}
\caption{Given are numerical values of the integral $I(m_P^2)$
(Eq. \ref{eq:aa}) in
GeV$^3$ for various values of the cutoff $\Lambda$ given in MeV. For
comparison, dimensional regularization of this integral coresponds to
the values 0.00246, 0.1213 and 0.1815 for $\pi$, $K$ and $\eta$
respectively.}
\end{center}
\end{table}

In Table 2, we
show the loop correction to the baryon masses for various values of
$\Lambda $. These are much smaller than the corresponding results in
dimensional regularization. In addition, since these results contain
an overall shift in $M_0$, one also notes that the SU(3) breaking is
decreased greatly, yielding very reasonable amounts of SU(3)
breaking. For any of these values of $\Lambda$, we can obtain an
excellent fit to the baryon masses. There is not much dynamical
content in such a fit, as the loop effects are now small enough that
we cannot demonstrate the presence of them in the data.
However, we see an excellent description without the need
for yet higher orders. The convergence of the expansion is now much improved.

\begin{table}
\begin{center}
\begin{tabular}{c|c|c|c|c|c}
 &dim. &$\Lambda=300$&$\Lambda=400$&$\Lambda=500$&$\Lambda=600$\\
\hline\\
$N$&-0.31&-0.04&-0.11&-0.22&-0.40\\
$\Sigma$&-0.67&-0.03&-0.08&-0.18&-0.34\\
$\Lambda$&-0.66&-0.03&-0.08&-0.18&-0.34\\
$\Xi$&-1.02&-0.02&-0.06&-0.15&-0.29
\end{tabular}
\caption{Given (in GeV) are the nonanalytic contributions to baryon
masses in dimensional regularization and for various values of the
cutoff parameter $\Lambda$ in MeV.}
\end{center}
\end{table}

\section{Summary}
   The procedure described above matches well with the goals of
effective field theory because it keeps only the long-distance portion
of loop diagrams. It turns out that only a portion of kaon and eta
loops are truly long-distance, and that inclusion of just
these effects
yields an chiral expansion that is well behaved.

We now understand the origin of the previous problem with the
convergence of the chiral expansion in baryons. The previous
analyses using
dimensional regularization had implicitly included spurious
short-distance physics in the loop calculation. This required
correction at higher orders
in the energy expansion, and appeared to lead to a poor convergence.
The cutoff regularization excludes this spurious physics, and
therefore leads to a better description.

It is certainly true that it is more difficult, and occasionally
more subtle \cite{7,8}, to work with a momentum space cutoff than with the usual
dimensional procedure. Indeed in the meson sector the dimensional formalism
works fine. However, in baryons the improved convergence properties obtained
by use of a cutoff justifies the
extra effort. Indeed, this development may finally allow realistic
phenomenology to be accomplished in SU(3) baryon chiral perturbation
theory.
\section*{Acknowledgments}
This research was supported in part by the National Science Foundation.

\end{document}